\documentclass[fleqn,usenatbib]{mnras}

\usepackage[T1]{fontenc}

\DeclareRobustCommand{\VAN}[3]{#2}
\let\VANthebibliography\thebibliography
\def\thebibliography{\DeclareRobustCommand{\VAN}[3]{##3}\VANthebibliography}

\usepackage{graphicx}	% Including figure files
\usepackage{amsmath}	% Advanced maths commands
\usepackage{amssymb}	% Extra maths symbols

\title[Host properties and environment of AGNs]{On the relation of host properties and environment of AGN galaxies across the standard optical diagnostic diagram}

\author[Perez et al.]{
Noelia R. Perez,\thanks{E-mail: noeliarocioperez@gmail.com (NP)}
and Georgina Coldwell,\thanks{E-mail: georginacoldwell@gmail.com (GC)}
\\
Facultad de Ciencias Exactas, F\'{i}sicas y Naturales, Departamento de Geof\'{i}sica y Astronom\'{i}a, CONICET Universidad Nacional
de San Juan, Av. Ignacio de la\\  Roza 590 (O), J5402DCS, Rivadavia, San Juan, Argentina
}
\date{Accepted XXX. Received YYY; in original form ZZZ}

\pubyear{2021}

\begin{document}
\label{firstpage}
\pagerange{\pageref{firstpage}--\pageref{lastpage}}
\maketitle

\begin{abstract}
We study the host properties and environment of active galactic nuclei (AGNs) galaxies, taken from SDSS-DR12, across the $\text{[O III]}/\text{H}\beta$ vs $\text{[N II]}/\text{H}\alpha$ diagnostic diagram. We select AGN subsamples defined as parallel and perpendicular to the star-forming locus on the BPT diagram based on the Kauffmann et al. and Schawinski et al. criteria.
For parallel subsamples we find that AGN host properties exhibit a morphological evolution as they become more distant to the star-forming sequence. The local density environment shows a more evident morphology-density relationship for subsamples mainly formed by Composite and Spiral galaxies than those containing LINERs and Seyferts, where the AGN emission is the dominant source. 
We also analyse the properties of the five closest AGN neighbours observing no significant differences in the environment, although the AGN host properties of every subsample have noticeable variations.
The AGNs belonging to perpendicular subsamples show clear differences on their host properties from left top to right bottom on the diagram. However, the analysis of the local density environment do not reflect strong dependency with the host AGN properties. This result is reinforced by the characteristics of the AGN neighbouring galaxies.
These findings suggest that mixed AGN/star-forming galaxies present environmental features more similar to that of non-active galaxies. However, as AGNs at the centre of the more evolved galaxies become the dominant source, the environment tends to provide suitable conditions for the central black hole feeding with an increasing content of gas and likelihood of a higher merger rate.

\end{abstract}


%%%%%%%%%%%%%%%%%%%%%%%%%%%%%%%%%%%%%%%%%%%%%%%%%%

%%%%%%%%%%%%%%%%% BODY OF PAPER %%%%%%%%%%%%%%%%%%

\section{Introduction}

Galaxy spectra provide relevant information about the physical processes taking place within galaxies. 
Numerous galaxy properties such as stellar age population, star formation rate (SFR), chemical abundance, nuclear activity, etc. can be obtained from the optical spectrum. 
Further, the existence of emission lines as an outstanding spectral feature allows to distinguish between the different mechanisms related with the energy sources.

In particular, the narrow emission lines in galaxies can be originated by different processes, therefore considerable effort has been made to be able to classify these objects using optical wavelengths information. 
\citet{Baldwin1981} were the first to propose the use of diagnostic diagrams considering line intensity ratios, as i.e. $\text{[O III]} / \text{H} \beta$  vs $\text{[N II]} / \text{H} \alpha$, known as BPT diagram, to distinguish emission-line galaxies according to their main excitation mechanism: photoionization by a power-law continuum source, photoionization by O and B stars, shock-wave heating or planetary nebulae.
Later, \citet{Veilleux1987} extended the classification scheme and defined the  well-know standard optical diagnostic diagrams, including the BPT diagram, $\text{[O III]} / \text{H} \beta$  vs $\text{[S II]} / \text{H} \alpha$ and $\text{[O III]} / \text{H} \beta$  vs $\text{[O I]} / \text{H} \alpha$, known as VO87 diagrams. In addition, the authors roughly estimated quantitative limits between H II region-like objects and narrow-line active galactic nuclei (AGNs).

Taking into account these earlier works, \citet{Kewley2001} used a combination of stellar population synthesis model, photoionization and shock models to determine a theoretical maximum starburst line, defined for the three standards optical diagnostic diagrams, establishing a lower limit to the mixed AGN and starburst emission.
Later, \citet{Kauffmann2003b} (hereafter Ka03) introduced an empirical line considering the dispersion in the modelling and the two distinctive galaxy emission-line sequences on the BPT diagram, which enables to clearly distinguish between pure H II regions and galaxies dominated by an AGN. Then,
objects lying in between Ka03 and \citet{Kewley2001} criteria are expected to present a combination of star formation and AGN emission in their optical spectra.

Regarding the properties of the narrow emission-line host galaxies, Ka03 performed a detailed analysis demonstrating that AGNs reside preferentially in massive and more concentrated galaxies. Moreover \citet{Kewley2006}, using an exhaustive AGN classification scheme, found that LINERs are older, more massive, and less concentrated than Seyfert or Composite galaxies.
Recently, \citet{Zewdie2020} analysed a sample of emission-line galaxies classified spectroscopically using the BPT diagram, and with morphological distinction based on Galaxy Zoo \citep{Lintott2011}, and found a direct relation of the SFRs in galaxies according to the classification criteria using diagnostic diagrams. Thus, LINER galaxies present low values of SFR, with morphological features typical of elliptical galaxies. Instead of that, Seyfert and Composite galaxies showed similar bimodal SFR distributions, exhibiting morphologies consistent with disc-type objects.

In addition, galaxy properties are strongly related to the environment they are located in.
Morphology-density relation indicates that early-type galaxies are located in high-density environments, having low SFR values. On the contrary, late-type galaxies lie preferentially in field or low-density regions presenting high SFR \citep{Dressler1980,Balogh1997}.  
However, several studies have shown that AGN host galaxies do not seem to follow this relation. 
\citet{Miller2003} found that AGN fractions, residing in  environments with different densities, are independent of the host galaxy morphology. Also, 
\citet{Padilla2010} found that active galaxies in clusters are bluer than galaxies without nuclear activity. In particular, Seyfert 2 galaxies with high and low accretion rate can be found residing in bluer and lower density regions than a well-defined control sample \citep{Coldwell2014}.
Also, LINERs were found to populate low-density environments in spite of their morphology, which is typical of high-density regions such as rich galaxy clusters \citep{Coldwell2017}.

The neighbouring galaxies of AGNs also present specific characteristics related with the more suitable conditions for the central black hole feeding. 
Thus, several works have demonstrated that AGN environs have high content of gas being particularly populated by blue, disc, and star-forming galaxies, compared to those in the vicinity of non-active galaxies with similar host properties \citep{Coldwell2006, Coldwell2009, Coldwell2018}. 
These results are consistent with a scenario where the higher merger rate and the lack of a dense intergalactic medium favour a higher occurrence of the different types of AGNs. 

By considering the particular conditions needed to support the existence of nuclear activity in galaxies, in this paper we analyse the AGN host galaxy properties from two different perspective with respect to their position on the BPT diagnostic diagram. Also, we study the surface local density environments and the characteristics of the close neighbouring galaxies.
 This paper is organized as follows: Section 2  gives a brief overview of the data and describes the sample selection of AGN emission-line galaxies according to their position on the BPT diagram.  In Section 3, we analyse the 
host properties and the environment of AGN subsamples using the local surface density and the features of the close neighbours. 
Finally, the main results are summarized and discussed in Section 4. Throughout the paper, we have assumed a cosmological model with $H_0 = 100 \ km \ s^{-1} \ Mpc^{-1}$, $\Omega_{m0} = 0.3$, and $\Omega_{\Lambda 0} = 0.7$.

\section{Data}

The analysis of this paper has been based on galaxy samples taken from the spectroscopic Data Release 12 of Sloan Digital Sky Survey\footnote{https://www.sdss3.org/dr12/} \citep[SDSS-DR12]{Alam2015}. 
This survey covers 14 555 square degrees of sky and includes imaging in five broad-bands (ugriz), reduced and calibrated using the final set of SDSS pipelines.
Additionally, the survey contains  galaxy and quasar spectra from the SDSS-III Baryon Oscillation Spectroscopic Survey \citep[BOSS,][]{Dawson13}. 

\subsection{Sample selection}

All catalogue data were obtained through $\rm SQL$ queries in $\rm CasJobs$\footnote{http://skyserver.sdss.org/casjobs/}. 
We select galaxies with spectroscopic information and extinction and k-corrected model magnitudes.
The catalogue was built using the publicly available emission-line fluxes corrected for optical reddening, employing the Balmer decrement and the \cite{calzetti00} dust curve. 
For this work, we assume an $R_V=A_V/E(B-V)=3.1$ and an intrinsic Balmer decrement $(H\alpha/H\beta)_{0}=3.1$ \citep{OM89}. Details about the line measurements are described by \cite{tremonti04} and \cite{brinch04}.

The emission-line galaxy sample was constructed with galaxies having $S/N > 2$ for all the lines involved in the BPT diagram ([N II]/$H\alpha$ vs. [O III]/$H\beta$) in agreement with 
 \cite{Juneau2014}. Additionally, we restrict to have $S/N > 2$ to the emission lines intervening in the VO87 diagrams ($[S II]/H\alpha$ and $[O I]/H\alpha$) providing a more reliable AGN classification \citep{Agostino2019} with a moderate reduction in sample size.
 Moreover, following \citet{Coldwell2017}, the sample was restricted to have a redshift range of $0.04 < z < 0.1$ obtaining $139 153$ sources. 

\subsection{Emission-line galaxies}

\begin{figure}
	\includegraphics[width=\columnwidth]{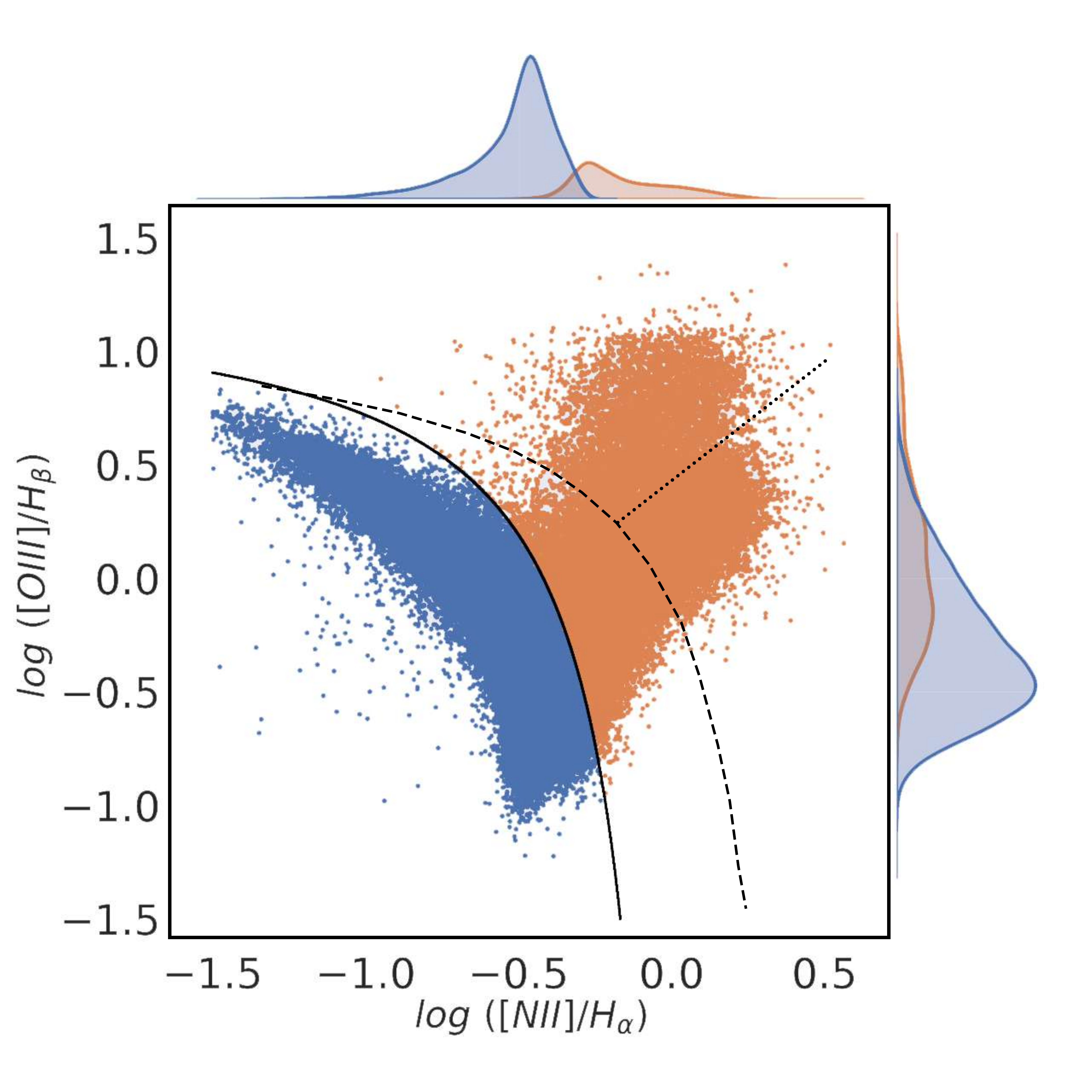}
    \caption{BPT diagnostic diagram. The solid line represents the empiric demarcation between HII galaxies and type 2 AGN defined by Ka03, the dashed line indicates the \citet{Kewley2001} theoretical demarcation, and the dotted line shows the empirical Seyfert-LINER separation determined by Sc07}. The histograms show the distributions of
    $\text{[N II]} / \text{H} \alpha$ and $\text{[O III]} / \text{H} \beta$ line ratios in logarithmic scale, on the x-axis and y-axis, respectively, for both samples. 
    \label{fig:fig1}
\end{figure}

In order to distinguish the different types of emission-line galaxies according to their main dominant energy source, such as gas photoionization by AGN \citep{Koski78,Groves04} or HII regions \citep{Huchra77,Kewley2001}, we use the BPT diagram as shown in Figure \ref{fig:fig1}, where the two populations can be clearly distinguished. In this figure the empiric curve defined by Ka03,
\begin{equation}
     \text{   log([O III] / }H\beta \text{ ) = 0.61 / [log([N II] / H}\alpha \text{ ) - 0.05] + 1.3},
\end{equation}
allows to separate the galaxies containing  H II regions from those with nuclear activity as the main responsible emission source. 
Then, galaxies lying above the Ka03 line are classified as AGN,  while star-forming galaxies are situated below this line, representing the $24.36\%$ and $75.64\%$ of the narrow emission-line galaxy sample, respectively.
Moreover, in this figure it is possible to observe the distributions of the emission-line ratios for both AGN and star-forming galaxies, where the $\text{[N II]} / \text{H} \alpha$ ratios differ significantly between the two samples although a substantial overlap is shown for $\text{[O III]} / \text{H} \beta$ ratios. 

Regarding to this classification criterion there is a risk of missing strongly obscured AGNs that may fall in the star-formation galaxy region of the diagram. However, besides the fact that the mean AGN fraction
in the local universe is approximately four times lower than that of star-forming galaxies, only a small fraction of these AGNs are heavily obscured \citep{Ricci2015} at low redshift.
On the other hand, it is possible to miss dust-enshrouded star-forming galaxies that do not show emission lines although this percentage is not significant in the local universe \citep{HG2013} concluding that the curve, defined by Ka03, allows a reliable emission-line galaxy classification.

 Type 2 AGNs  can be separated into  subclasses such as Composite, Seyfert, and LINER galaxies.  Thereby, \citet{Kewley2001} define  a  theoretical  discrimination, given by equation \ref{eq:kewley}, between galaxies with a mixed contribution of HII regions and AGN emission to the spectral features (Composite galaxies) from ``pure'' AGNs. Then Composite galaxies are located below this curve and above of the Ka03 line on the BPT diagram.
\begin{equation}
     \text{   log([O III] / }H\beta \text{) = 0.61 /  [log([N II] / H}\alpha \text{ ) - 0.47] + 1.19}\label{eq:kewley}
\end{equation}
Further, \citet{Schawinski2007} (hereafter Sc07) allow to classify the ``pure'' type 2 AGNs in Seyfert and LINER galaxies through equation \ref{eq:schawinski}.  Thus, Seyfert are placed above and LINER below the {Sc07} line  criterion, respectively.

\begin{equation}
     \text{   log([O III] / }H\beta \text{) = 1.05 log([N II] / H}\alpha \text{ ) + 0.45}\label{eq:schawinski}
\end{equation}

\subsection{Parallel and perpendicular classification of AGNs}

This work is aimed to investigate the dependence of the host properties of narrow emission-line galaxies, mainly dominated by an AGN,  with respect to their position on the BPT diagram and also the relationship with their environment.
Bearing this in mind, we define parallel and perpendicular subsamples, based on the Ka03 and Sc07 criteria, as it is described in the following sections.

\begin{figure}
	\includegraphics[width=\columnwidth]{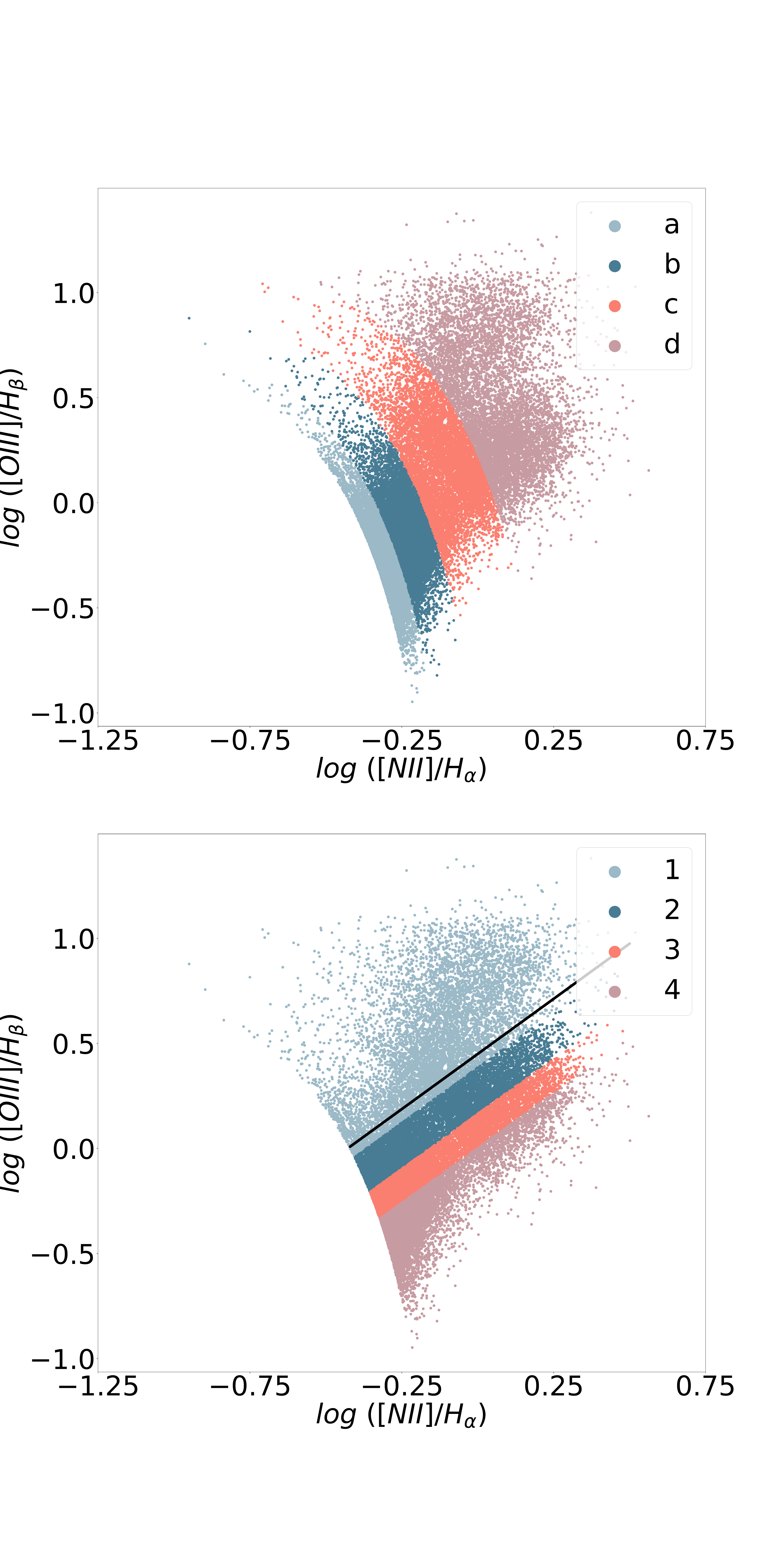}
    \caption{Top: BPT diagnostic diagram showing the subsamples defined parallel to the Ka03 curve criterion. Sample ``a'' is the closest to the Ka03 line and ``d'' is the farthest one. Bottom: BPT diagnostic diagram showing the subsamples defined perpendicular to the Ka03 criterion.  Subsample ``1'' is further to the left top of the diagram and subsample ``4'' is further to the right bottom. The solid black line indicates the empirical Seyfert-LINER separation determined by Sc07.}
    \label{fig:submuestras}
    
\end{figure}
\subsubsection{Parallel}
A curve line is shifted, parallel to the Ka03 line criterion defined in equation \ref{eq:kewley},  along both axes of the BPT diagram, as shown in Figure \ref{fig:submuestras} (top panel). We determine four bins of variable width, in dex in log line-ratio space, in order to obtain four subsamples with similar number of sources. These subsamples are named  as ``a'', ``b'', ``c'', and ``d'', where ``a'' is that closest to the  Ka03 curve and ``d'' is the set of galaxies farthest from the Ka03 criterion. The variable width of bins ``a'', ``b'', ``c'', and ``d'' corresponds to 0.06, 0.12, 0.19, and 0.62 dex, respectively.
The proportion of AGN subclasses, described in the previous section, for every defined parallel bin is quantified in Table \ref{tab:muestras}. As it is expected, bins  ``a'' and ``b'' are fully populated by Composite galaxies, decreasing to 35\% and 0\% for bins ``c'' and ``d'', respectively. LINERS are the predominant galaxies in bins ``c'' and ``d'', being $\sim13$ per cent and $\sim18$ per cent higher than Seyferts, respectively.

\subsubsection{Perpendicular}
We determine perpendicular subsamples, to the star-forming locus, taking as a reference the Sc07 line criterion, given by equation \ref{eq:schawinski}, defined to characterize the position of Seyfert and LINER galaxies on the BPT diagram. Hence, we consider the Sc07 line approximately perpendicular to the Ka03 criterion. Then, we shift this line from left top to right bottom in order to obtain four subsamples with similar number of sources in each bin of variable width, in dex in log line-ratio space. The subsamples are named ``1'', ``2'', ``3'', and ``4'', whose variable width of the bins corresponds to 0.76, 0.11, 0.08, and 0.30 dex, respectively. The  ``1'' is the subsample lying closest to the top of the diagram, as shown in Figure \ref{fig:submuestras} (bottom panel). Table \ref{tab:muestras} shows that bin ``1'' is mainly composed of Seyfert AGNs, while the proportion of Composite galaxies is dominant in the remaining bins. Further, the higher fraction of LINERs is observed for bin ``2''.

\begin{table}
	\centering
	\caption{Fraction of AGN subclasses and morphological classification in each subsample.}
	\label{tab:muestras}
	\begin{tabular}{cllllll} 
		\hline
		Sub- & LINER & Seyfert & Composite & E & S & U \\
		sample & (per cent) & (per cent) & (per cent)& (per cent) & (per cent) & (per cent) \\
		\hline
		a & 0.00 & 0.01 & 99.99 & 1.52 & 52.72 & 45.76 \\
		b & 0.00 & 0.36 & 99.64 & 2.24 & 52.57 & 45.19\\
		c & 38.63 & 25.94 & 35.43 & 6.64 & 40.14 & 53.22\\
		d & 58.93 & 40.93 & 0.14 & 11.09 & 33.94 & 54.97\\ 
				\hline
		1 & 8.05 & 67.17 & 24.78 & 3.55 & 39.94 & 56.51\\
		2 & 41.70 & 0.00 & 58.30 & 6.78 & 42.27 & 50.95\\
		3 & 31.44 & 0.00 & 68.56 & 6.07 & 46.32 & 47.41\\
		4 & 16.26 & 0.00 & 83.74 & 4.78 & 50.48 & 44.74\\
		\hline
	\end{tabular}
\end{table}

\subsection{Morphological classification}

In order to study the morphological features of active galaxies across the BPT diagram  we cross-correlated the AGNs from the defined subsamples with the Galaxy Zoo catalogue \citep{Lintott2011}. This catalogue comprises a morphological classification of nearly 900 000 galaxies, from the spectroscopic data of the SDSS, where hundreds of thousands of volunteers contributed to discriminate the galaxies giving the fraction of votes in six categories: elliptical, spiral, spiral clockwise, spiral anticlockwise, merger, or uncertain.
The vote fraction threshold, for the table with the galaxy classifications available for this project\footnote{https://data.galaxyzoo.org}, is 0.8.  Details about the classification bias are described in \cite{Bamford2009}.
For the AGN subsamples we calculate the percentage of galaxies classified as spiral (S), elliptical (E), or uncertain (U) as it is shown in Table \ref{tab:muestras}.

\section{Analysis}
Several meaningful physical properties of galaxies, such as stellar masses, emission-line fluxes, stellar age indicators, gas-phase metallicities, etc. have been measured for SDSS galaxies. 
The procedures adopted to estimate these parameters are described by \cite{brinch04}, \cite{tremonti04} and \cite{kauffmann2003a}. 
These data are available from the MPA-JHU\footnote{http://www.sdss.org/dr12/spectro/galaxy\_mpajhu/}.

In this research we explore the host galaxy properties of narrow emission-line galaxies, classified as AGN, and from their environment. In this context, we analyse the morphology, age, SFR and colours of galaxies as indicative parameters.
In particular, we use the break index $D_n$(4000) as a stellar age indicator \citep{kauffmann2003a}.
This parameter represents an important effect in the spectra of old stars, occurring at 4000 \AA, and arises by an accumulation of a large number of spectral lines in a narrow region of the spectrum, so that the oldest galaxies have higher $D_n$(4000) values than youngest ones.  
We adopted the \cite{Balogh1999} definition as the ratio of the average flux densities in the red continuum (4000-4100 \AA) with respect to that in the blue continuum (3850-3950 \AA). 
Further, we use the SFR parameter derived by modelling the emission lines for star-forming galaxies, and estimated from the integrated photometry using the correlation with $D_n$(4000) for Composite and AGN galaxies \citep{kauffmann2003a}, following the methods described in \cite{brinch04}. The fibre aperture corrections were done based on broad-band photometry \citep{Salim07}.

In addition, colours of galaxies can be used to discriminate their stellar population, evolution, and environment  \citep{Vulcani15}. 
Hence, in galaxy clusters, the larger fraction of red galaxies indicate an old stellar population within galaxies having a lower SFR. 
On the other hand, galaxies residing in poor groups and field present bluer colours and higher SFR.
To analyse the colour of galaxies, $M_g-M_r$, we use model magnitudes which are more appropriated for extended objects, such as galaxies, by providing more robust estimation of this parameter.  These magnitudes are k-corrected using the empirical k-corrections presented by \cite{Omill11}. 

To complement the analysis using the age and colour parameters we use the concentration index as an indicator of the morphology, following the works of \citet{Shimasaku2001} and \citet{Nakamura2003}.
The concentration index, $C$, is defined as the ratio of the radii containing 90 and 50 percent of the Petrosian r flux, respectively.
Hence, elliptical galaxies are expected to have higher concentration indices than the spiral ones \citep{Strateva2001}. 
Then, galaxies with a de-Vaucouleurs profile have values of $C \sim 3.3$ and disc type galaxies have a concentration index $C \sim 2.4$.

The host properties are able to reflect the evolutionary stage of galaxies belonging to the AGN and star-forming samples. 
Different works have shown that a bimodal colour distribution is observed in different galaxy environments \citep{Pandey2020}.
Furthermore, the relationship observed in the colour-mass  and colour-SFR diagrams are able to show two distinctly separate morphological populations with different evolutionary scenarios \citep{Schawinski2014} allowing to distinguish between red and old galaxies, corresponding with the more evolved sources, from the blue and young ones.

 \begin{figure}
	\includegraphics[width=\columnwidth]{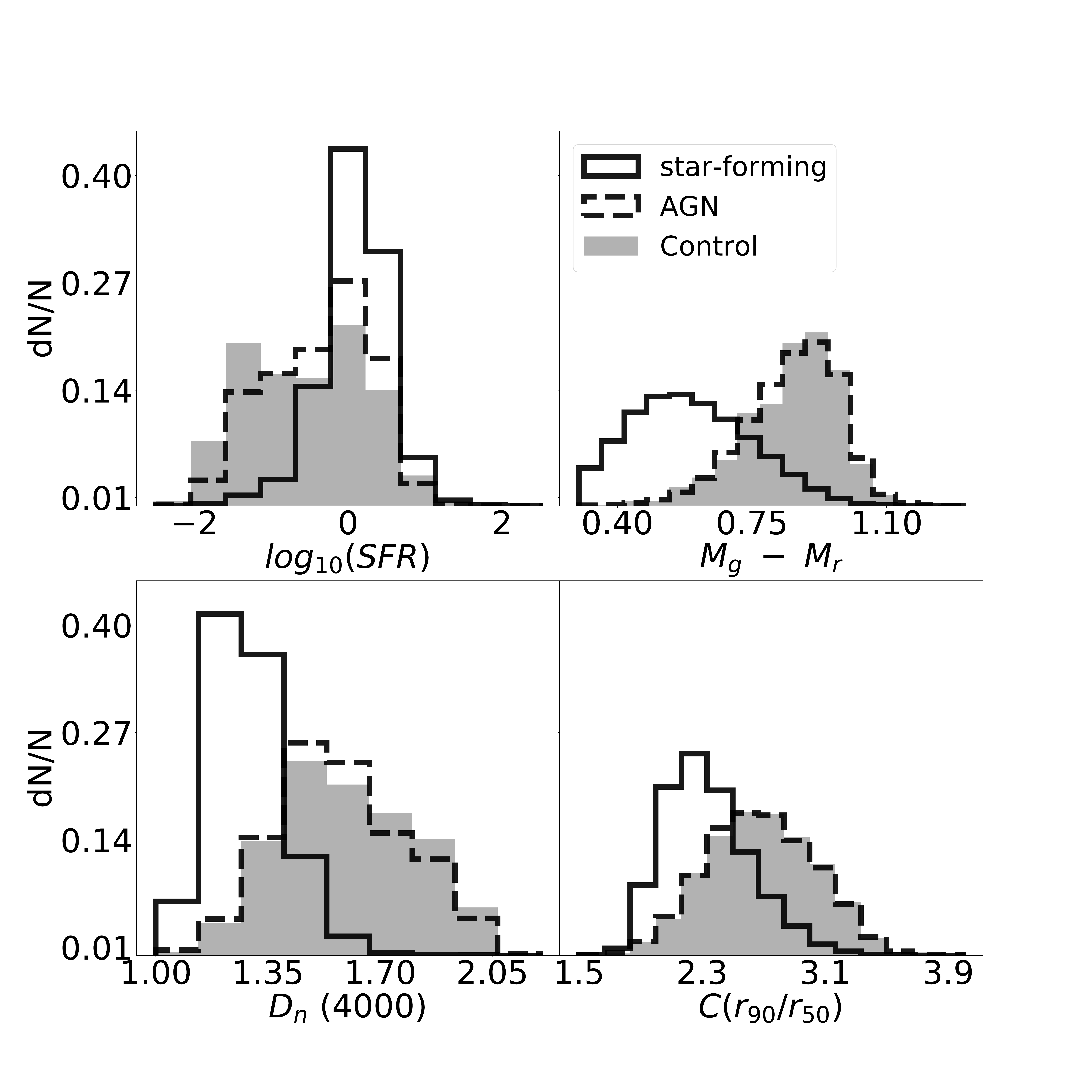}
    \caption{Normalized distributions of galaxy properties for emission-line galaxies classified as star-forming galaxies (solid line) and AGNs (dashed line). The shadow distributions correspond to the star-forming control sample selected to match the AGN host galaxy properties.}
    \label{fig:fig2}
\end{figure}

The host galaxy property distributions of the emission-line galaxy sample shown in  Fig. \ref{fig:fig2} reflect an evident distinction from their colours, age, and morphological features between objects classified as AGNs from the star-forming galaxies.
Particularly, from the  galaxy colours two Gaussian distributions can be distinguished where AGN host galaxies correspond to the red galaxy population and the star-forming galaxies to the blue ones.
Thus, AGN hosts have a bulge-type morphology, being redder and older than star-forming galaxies \citep{Kewley2006,Zewdie2020}.
In addition, the AGNs have lower values of the SFR parameter than the star-forming galaxies as expected.

The properties of the whole AGN sample are well known; however, how they change with respect to their position on the BPT diagram is very interesting to study. In the following sections, we analyse the host properties for AGNs from the defined parallel and perpendicular subsamples.

\subsection{Colour and SFR distributions}\label{sec:sfr_colour}

\begin{figure}
	\includegraphics[width=\columnwidth]{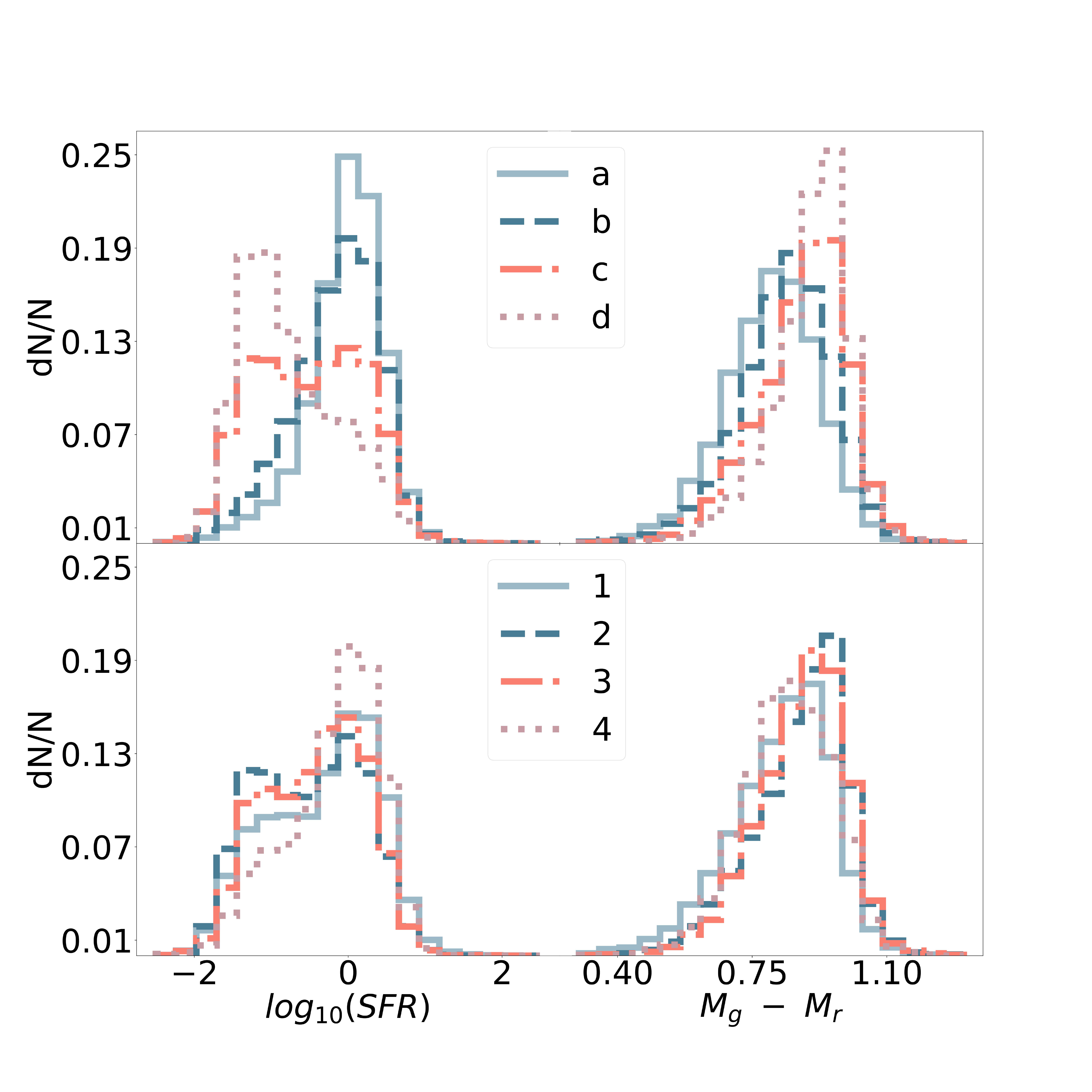}
    \caption{Normalized distributions of SFR, in logarithmic scale, and colours ($M_g-M_r$) for AGNs from parallel (top) and perpendicular subsamples (bottom).}
    \label{fig:sfr_gr_prop}
\end{figure}

The normalized distributions of $M_g-M_r$ colour and SFR are shown, in Figure \ref{fig:sfr_gr_prop}, for host galaxies belonging to parallel (top panels) and perpendicular subsamples (bottom panels), respectively.
In this figure, we observe that galaxies belonging to ``a'' subsample are bluer than galaxies from ``d'' bin. 
Galaxies from subsamples ``b'' and ``c'' present intermediate characteristics between those from ``a'' and ``d'' bins.
Further, the SFR parameter show a clear tendency. The distributions for galaxies from bins ``a'' and ``b'' are pretty similar, with higher values of SFR, since both samples are mainly formed by Composite galaxies. The galaxies in bin ``c'' show a bimodal distribution, while those belonging to bin ``d'' have the lowest values of SFR, suggesting a decreasing star formation activity as they drift away from the Ka03 criterion.

On the other hand, Figure \ref{fig:sfr_gr_prop} (bottom panels) shows the normalized property distributions for galaxies from perpendicular subsamples. 
AGNs from subsample  ``4'' show the highest values of SFR, even higher than those from subsample ``1'' (which are the bluest ones).
 Active galaxies belonging to bins ``2'' and ``3'' are similar, presenting lower values of SFR. This fact is reflected in the colours of galaxies from each bin. This result can be interpreted in terms of the content of AGN subclasses and their respective morphology for each bin. Thus, from Table \ref{tab:muestras} it is observed that subsample  ``4'' has the highest proportion of Composite and Spiral galaxies regarding to the remaining subsamples.

\subsection{Stellar age and morphological evolution} \label{sec:dn_c}

\begin{figure}
	\includegraphics[width=\columnwidth]{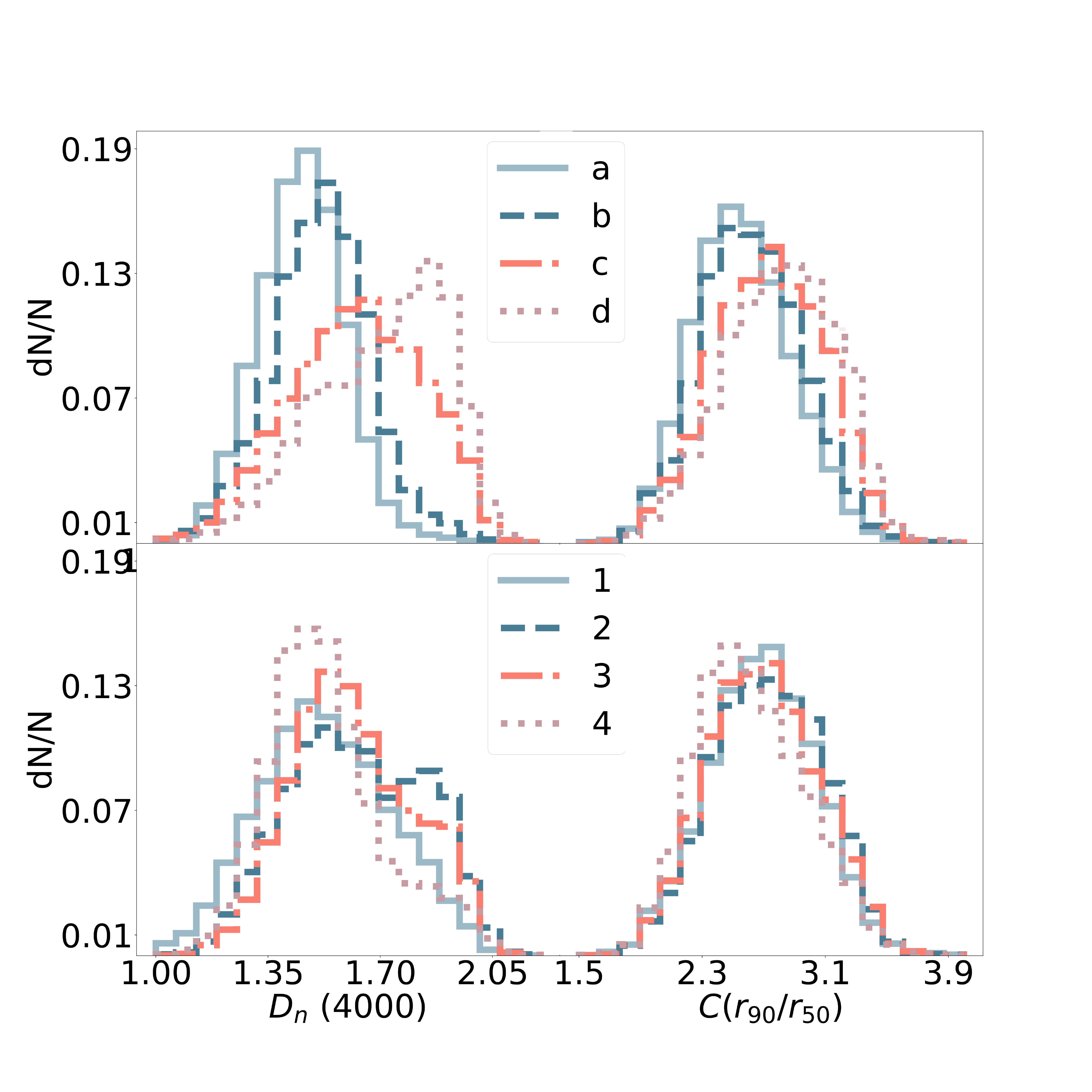}
    \caption{Normalized distributions of $D_n(4000)$ index and concentration index, C,  for AGN from parallel (top) and perpendicular subsamples(bottom).}
    \label{fig:dn_c_prop}
\end{figure}

For parallel subsamples, the distributions from Figure \ref{fig:dn_c_prop} (top panels) show that the concentration index, $C$, points out that galaxies closer to the Ka03 line are less compact than the farthest ones.
Also, it is important to remark that the $D_n(4000)$ parameter, used as a good estimator of the stellar age, presents the more evident distinction between the subsamples ``a'' and ``d''.
In all cases, galaxies from subsamples ``b'' and ``c'' present intermediate characteristics.
From Table \ref{tab:muestras} it is possible to observe the proportion of AGN subclasses and morphology reinforcing this finding. From subsample ``a'' to ``d'' the percentage of composite galaxies decrease sharply, increasing the amount of ``pure'' AGNs. Also, 
the fraction of spiral galaxies decrease, while the proportion of elliptical increase.
These results indicate the existence of a morphological evolution of the AGN host galaxies considering perpendicular distances from the tight sequence of star-forming galaxies, defined in parallel bins to the K03 line criterion. 

The property distributions for active galaxies belonging to perpendicular bins, shown in Figure \ref{fig:dn_c_prop} (bottom panels), allow to observe that AGNs from subsample ``1'' have the lowest values of 
$D_n(4000)$. Subsamples ``2'' and ``3'' have distributions corresponding to galaxies older than those of subsample ``1''. A noticeable bimodality is shown in subsample ``2'', indicating the presence of two distinctive galaxy populations in comparison with the remaining subsamples.
Galaxies belonging to subsample “4” have a transitional behaviour with respect to the age of the other subsamples. 
The concentration index distributions for all the subsamples are comparable with the exception of those active galaxies from subsample ``4'' which are slightly the less compact, in agreement with the proportion of Composite and Spiral galaxies in this bin.

In order to statistically quantify the  observed discrepancies between the property distributions of the subsamples, we applied a goodness-of-fit Kolmogorov-Smirnov test (KS) to the data. In all cases, we obtained $p < 0.05$ rejecting the null hypothesis that the samples were drawn from the same distributions.

\subsection{AGN host properties vs. environment}

Taking into account the meaningful contrast in the host galaxy properties, it is expected to be reflected in the environment. 
The well-known morphology-density relation establishes that  evolved  elliptical galaxies  are  found  inhabiting  high-density  environments while blue disc galaxies commonly reside in low density regions \citep{Dressler1980,Dominguez01,Cooper2010}.
However, several works \citep{Popesso06,Padilla2010,Coldwell2009} have found that AGNs do not follow the expected morphology-density relation of galaxies without nuclear activity. 
This effect is very noticeable even when AGN subsamples are studied separately as the case of Seyfert 2 \citep{Coldwell2014} and LINER galaxies \citep{Coldwell2017,Coldwell2018}.

A simple method to study the performance of the morphology-density relationship in our AGN sample is throughout the comparison with a well-determined control sample, selected according to the work of \citet{Perez2009}. 
Then, in a similar way to \cite{Coldwell2017,Coldwell2018}, we build a control sample using a Monte Carlo algorithm that selects sources from the star-forming galaxy sample with similar  distributions of redshift, colour, age, and morphology than that from the AGN sample.The KS test yields that the null hypothesis is true at a 90 per cent confidence level for all these parameters. Hence, the AGN and the control samples can be reliably considered drawn from the same distribution.
The control sample is shown in the shadow histograms from Figure \ref{fig:fig2} matching the $Mg-Mr$, $D_n(4000)$ and $C$ index AGN properties.  In Fig. \ref{fig:fig2}, we also show the SFR parameter, although it is important to notice that for the Control sample it was not forced to match that corresponding to AGN sample.
 
\begin{figure}
	\includegraphics[width=\columnwidth]{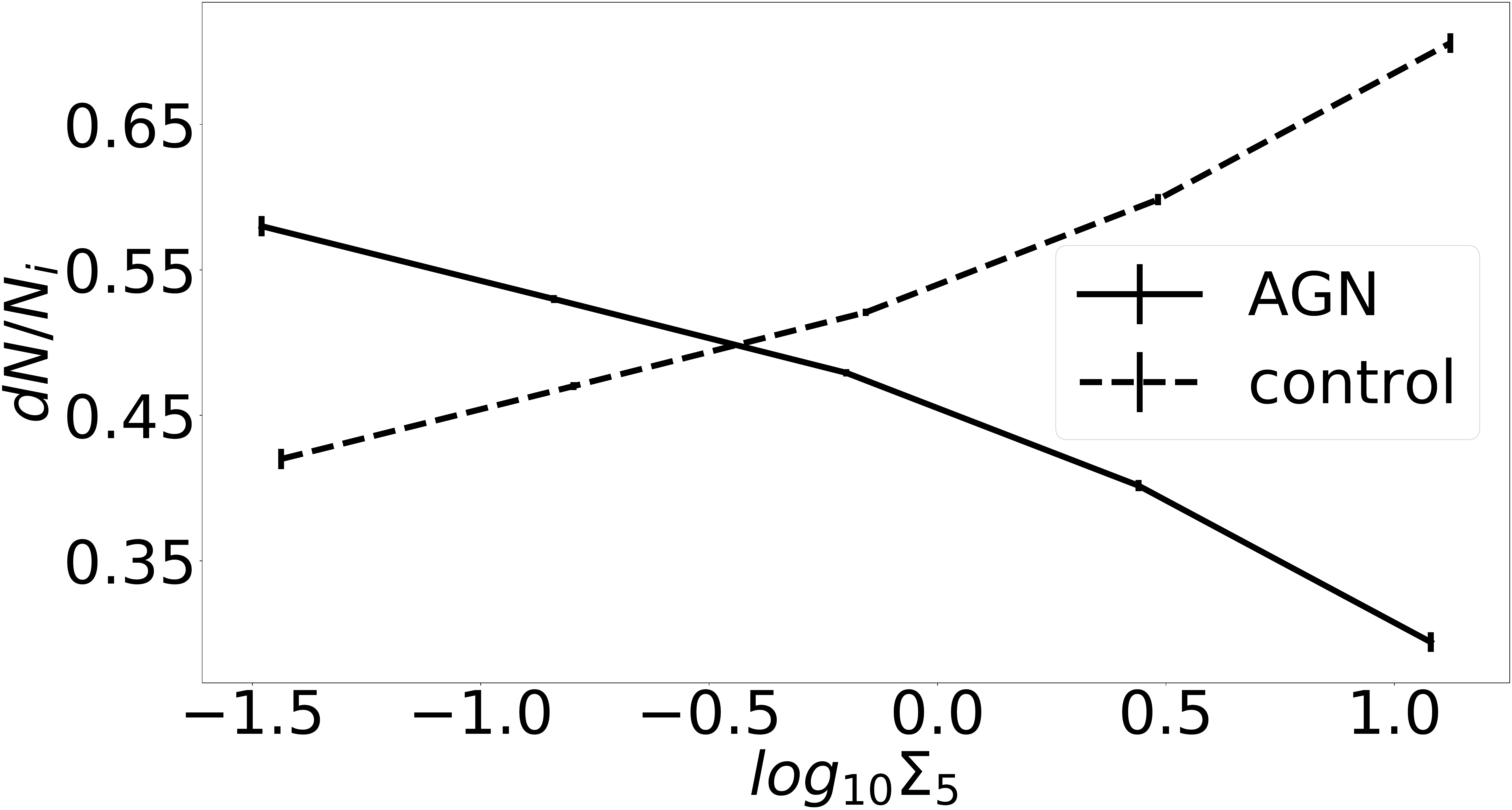}
    \caption{Fraction of AGNs (solid line) and star-forming galaxies from the control sample (dashed line) as a function of the surface density in logarithmic scale, $\Sigma_5$.}
    \label{fig:fig3}
\end{figure}

To perform the study of the density environment, we use the local galaxy density estimator given by $\Sigma_5 = \frac{5}{\pi d_5^2}$, where $d_5$ corresponds to the projected distance of the fifth neighbour brighter than $M_r < -20.5$ and with a radial velocity difference $\Delta V = 1000 \ km \ s^{-1}$ to include galaxies within systems with large velocity dispersion \citep{balogh2004}.
This  method considers a variable scale estimator using a systematically larger scale in lower density regions, which  improves  sensitivity  and  precision  at  low  densities. 
In addition, this  two-dimensional density estimator uses the redshift information to reduce the projection effects.
 
Bearing in mind that the control sample was built from the star-forming galaxy sample in order to reproduce the galaxy property distributions of AGNs, and so, they could be expected to reside in similar density environments.
Figure \ref{fig:fig3} shows the fraction of AGN and control samples with respect to the total number of objects for a given bin of the local density, $\Sigma_5$, in logarithmic scale.
From this figure it is possible to observe that although both samples have matched property distributions corresponding to red, old, and bulge-type galaxies, at the highest density environment the fraction of AGNs drops sharply approximately to the half of the control sample fraction.
This trend is reversed for lower density environment where the fraction of AGNs is, at least, 20\% higher than the fraction of galaxies from the control sample.
This result, in agreement with previous works \citep{Padilla2010,Coldwell2017,Coldwell2018}, reinforces the idea that the low-density environments, where there is a high probability of merger rate and a sufficient gas reservoir, provide suitable conditions for a higher AGN occurrence.
The error bars in this figure and in all figures of this paper were calculated using bootstrap error resampling technique \citep{Barrow1984}.

The results of the previous subsection suggest the existence of a dependence of the AGN host properties and its position on the BPT diagram according to the schema defined to select the subsamples analysed in this paper.
The environmental analysis of these galaxies could give light to the potential physical processes that may favour the occurrence of AGNs in every subsample.
For the sake of consistency, the study of the environment is performed throughout the use of the local density estimator $\Sigma_5$ parameter in order to obtain reliable estimations of the results for all the subsamples.

\subsubsection{Parallel subsamples}

\begin{figure}
	\includegraphics[width=\columnwidth]{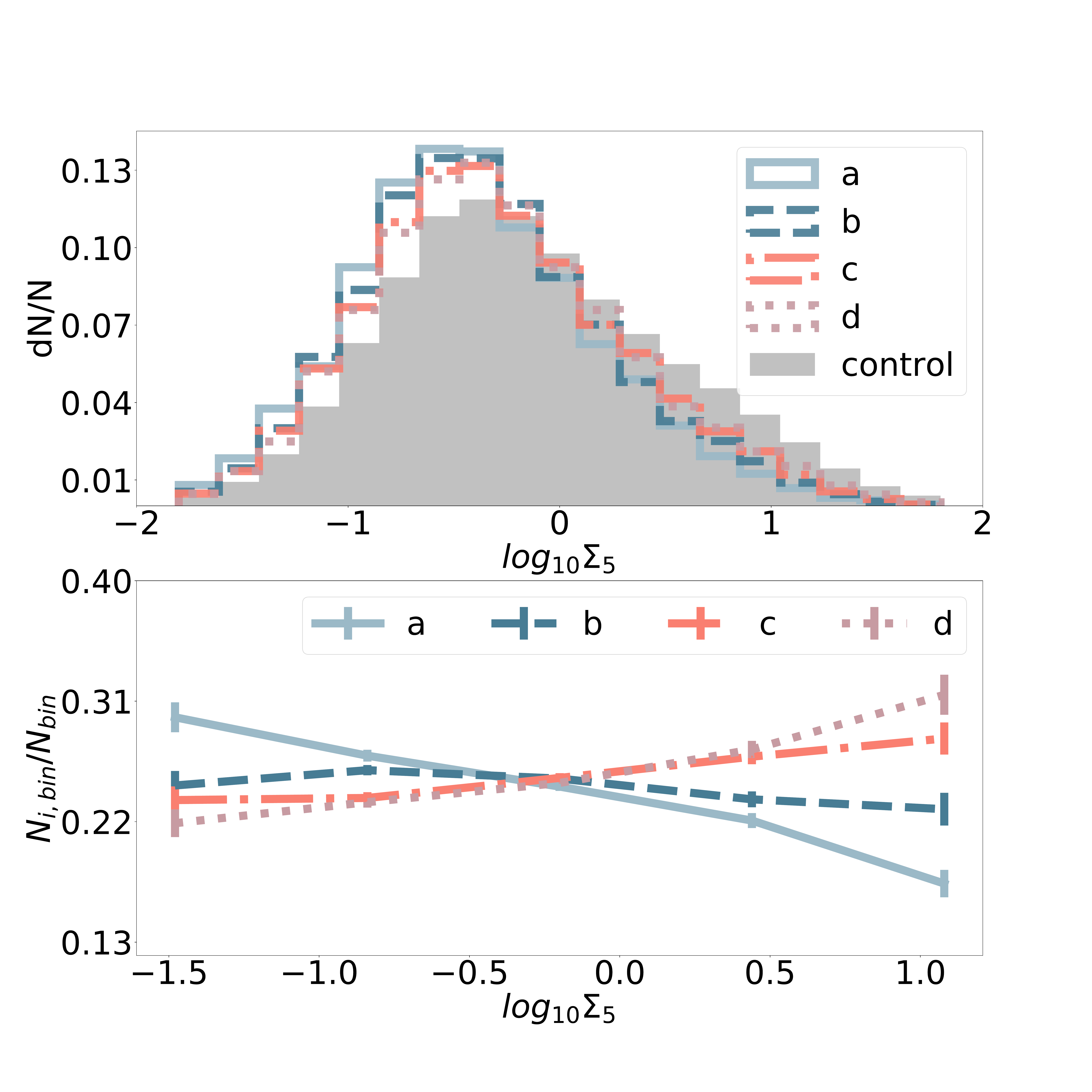}
    \caption{Top: Normalized distributions of the surface density in logarithmic scale, $\Sigma_5$, for the AGN subsamples and control sample (shadow).
    Bottom: Fraction of AGNs as a function of $\Sigma_5$. The solid line corresponds to ``a'' subsample, dashed line to ``b'', 
    dash-dotted to ``c'', and dotted line to ``d''  subsample, respectively.}
    \label{fig:par_Hsigma5}
\end{figure}

To explore the density environment of galaxies in every subsample, their $\Sigma_5$ distributions are shown in Fig. \ref{fig:par_Hsigma5} (top panel). For comparison, the local density distribution of the control sample, built matching the host galaxy properties from the entire AGN sample, is also included. Then, although all the AGN subsamples have lower density environment than the control sample, as it could be expected from Fig. \ref{fig:fig3}, slight differences can be perceived.

In order to quantify these small variations between the AGN subsamples the fraction of galaxies belonging to every parallel bin, with respect to the total number of objects for a given bin of $\Sigma_5$, is drawn in Fig. \ref{fig:par_Hsigma5} (bottom panel).
Galaxies in subsample ``a'' show a clear morphology-density relation consistent with their host properties, given that these galaxies are the youngest and bluest from the whole AGN sample, as it is described in Sects. \ref{sec:sfr_colour} and \ref{sec:dn_c}. 
Then, for subsample ``a'', the $\sim30$\% of galaxies belong to the lowest
density environment, representing an excess of $\sim6$\% respect to galaxies in the remaining parallel subsamples. This fraction decreases up to $\sim 20\%$ for the highest local density.

The tendencies shown by subsamples ``b'', ``c'', and ``d'' do not follow strictly the morphology-density relation as it could be expected from their host properties. 

Instead of that, the three subsamples have similar fraction of galaxies ($\sim$ 24\%) at lower values of  $\Sigma_5$, within the errorbars. 
Moreover, on the opposite extreme where $log_{10}\Sigma_5$ $> 1.0$, the three subsamples present a moderate difference, within 2 $\sigma$ errors, where the subsample farthest from the Ka03 criterion presents the  highest fraction of galaxies.
Thus, although galaxies belonging to  ``b'', ``c'', and ``d'' bins have very distinctive host properties, their local density environment are pretty similar except for the highest $\Sigma_5$ values.
Due to the position on BPT diagram, the subsample ``a'' is  fully formed by Composite galaxies, with more than the 50\% classified as Spiral, as can be appreciated in Table \ref{tab:muestras}. Thus, this fact could explain the observed trend in the morphology-density relation. However, although subsample ``b'' have comparable proportion of Composite and Spiral galaxies than ``a'', the host properties differ somewhat between both subsamples which is reflected in the trend shown in Fig. \ref{fig:par_Hsigma5} (bottom panel). For subsamples ``c'' and ``d'', the fraction of spiral galaxies is reduced, while the proportion of ``pure'' AGNs with lower SFR values grow and the relation with the environment become less sensitive to the density.
This result is in agreement with \citet{Miller2003}, who found that SFR fractions decrease for denser environments and AGN fractions remain constant in environments with different densities, such as clusters and voids.

Further, galaxy colours provide an indirect constrain on the evolutionary history of galaxies. 
So, in high-density regions, such as groups or galaxy clusters
, a large fraction of red galaxies indicating an older population with low star formation are expected to be found. 
On the contrary, galaxies in poor groups or in the field are bluer, presenting stronger SFR typical of young sources \citep{Dominguez01}.
In addition, the colours have been found to be the most predictive parameter indicative of environment, both for galaxies in the field \citep{Blanton2005} and in groups \citep{Martinez2006}.
In this section, we also quantify the dependence of the more evolved galaxies with respect to the density environment based on previous works \citep{Coldwell2009,Coldwell2014,Coldwell2018} in order to study their behaviour for our defined subsamples. Hence, we use galaxy colours and stellar age indicator to estimate the fraction of host galaxies redder and older than the mean values of these parameters for the main galaxy sample in the redshift range considered in this paper. We calculate the fraction of galaxies, in every subsample, redder than  $M_g - M_r > 0.76$ and older than $D_n(4000) > 1.53$, as a function of the local density. 

\begin{figure}
	\includegraphics[width=\columnwidth]{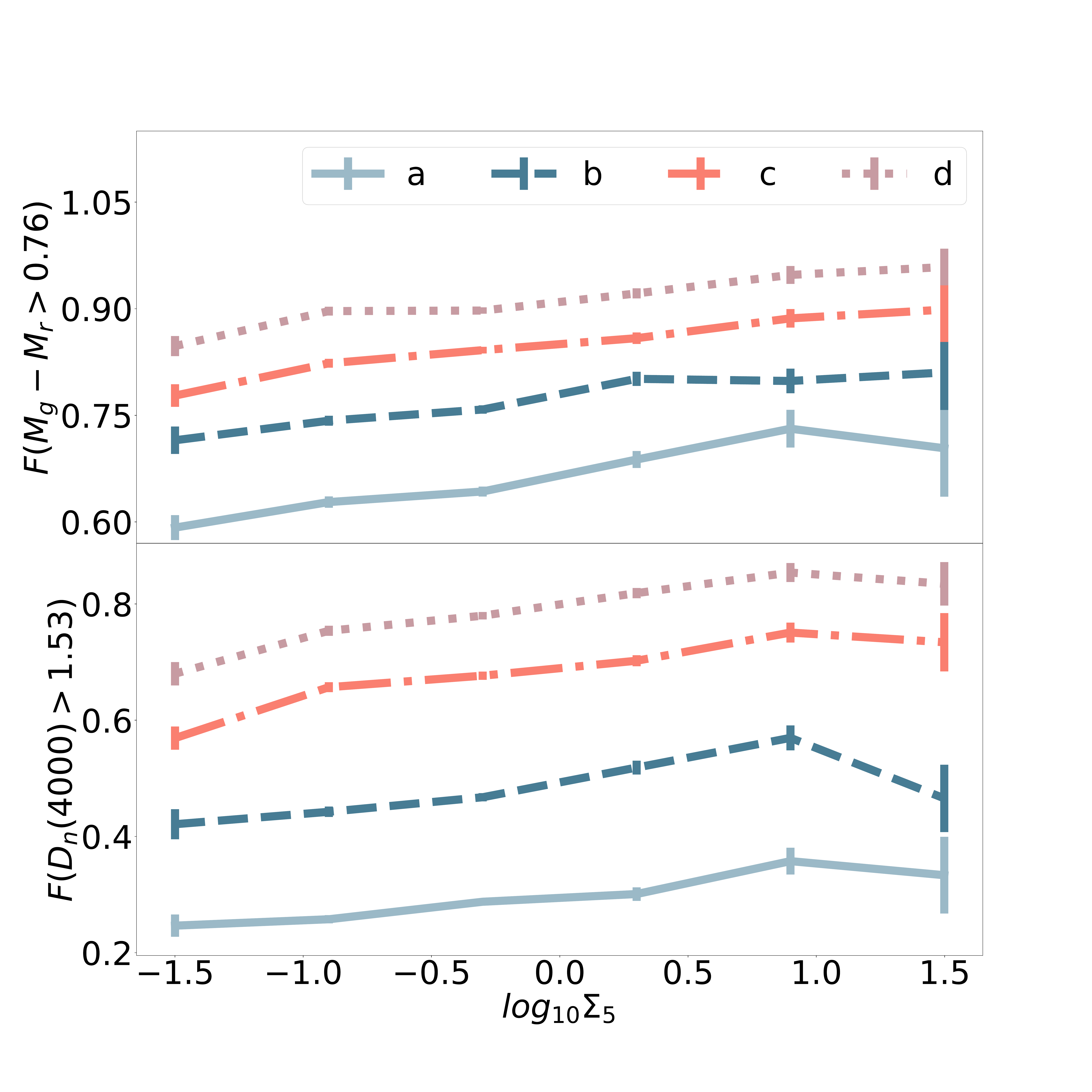}
    \caption{Fraction of red ($M_g - M_r > 0.76$) and old ($D_n(4000) > 1.53$) galaxies, as a function of the local density, for every parallel subsample. The solid line corresponds to ``a'', dashed line to ``b'', dash-dotted to ``c'', and dotted line to ``d'' subsamples, respectively.}
    \label{fig:par_F-S5}
\end{figure}

Figure \ref{fig:par_F-S5} shows that the fraction of bulge-type galaxies, redder and older than the mean galaxy population, increases significantly from bin ``a'' to ``d'' as they move away from the Ka03 line criterion, in agreement with the host galaxy distributions observed in Figs. \ref{fig:sfr_gr_prop} and \ref{fig:dn_c_prop}. 
With respect to the galaxy colours, the ``a'' subsample contains approximately the $65$per cent of these red galaxies, and the subsample ``d'' more than the $85$ per cent. Further, while the ``a'' subsample is formed, on average, by the $33$ per cent of galaxies older than the mean galaxy population, more than the $70$ per cent of galaxies from bin ``d' are older than the mean. 
The subsamples ``b'' and ``c'' present intermediate percentages for both parameters.
It is important to notice that the fraction of redder galaxies increases slightly from low to high local density environments, being more evident for subsample ``a''.
Moreover, the fractions of older galaxies present similar tendencies being still less sensitive to the density environment.

\begin{figure}
	\includegraphics[width=\columnwidth]{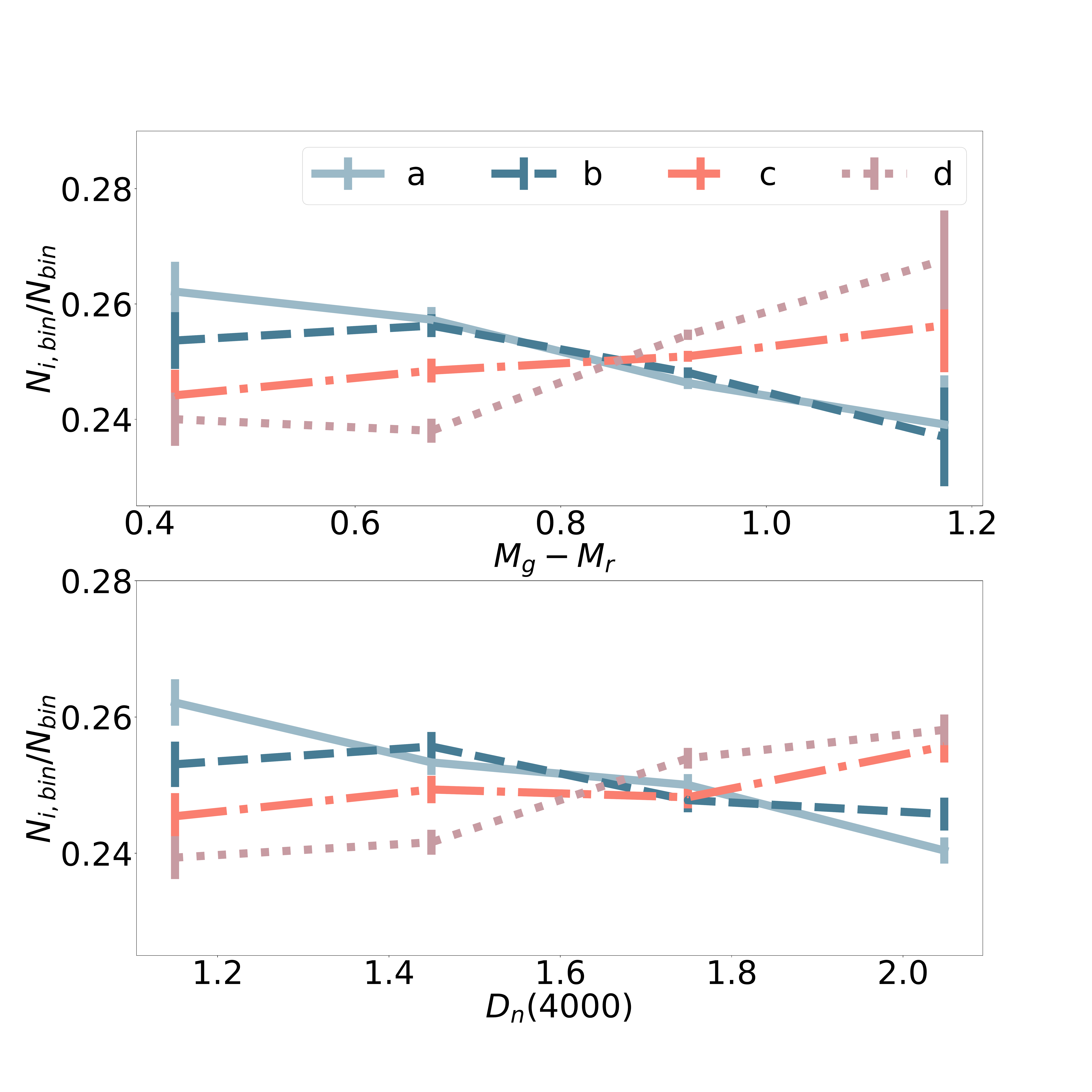}
    \caption{Fraction of the five closest neighbouring galaxies from AGNs as a function of the
    $M_g - M_r$ colour and the $D_n(4000)$ stellar age indicator. The solid line corresponds to ``a'', dashed line to ``b'', 
    dash-dotted to ``c'', and dotted line to ``d'' subsamples, respectively.
}
    \label{fig:par_H5tavecina}
\end{figure}

The characteristics of galaxies are strongly related with the environment. 
Thus, it is expected that the neighbouring galaxies of the subsamples have similar features to their corresponding central host galaxies.
In this line, the properties of the five closest neighbouring galaxies brighter than $M_r=-20.5$, used to estimate $\Sigma_5$, are analysed in order to understand their connection with the AGNs from every subsample.
It is observed, in Fig. \ref{fig:par_H5tavecina}, that the surrounding five closest galaxies of all the parallel subsamples present similar characteristics with small variations from the average fraction ($\sim 25$ per cent) for both colours and stellar ages.
The most noticeable signal is shown by the neighbouring galaxies from the AGNs belonging to bin ``a'', where the fraction of blue colours is almost 3$\sigma$ higher than the colours of galaxies in the vicinity of bins ``c'' and ``d'', although this difference becomes less significant for the fraction of red neighbours considering the errorbars. 
This result is comparable to that observed for the fractions of galaxies respect to the $D_n(4000)$ parameter.

\subsubsection{Perpendicular subsamples}

\begin{figure}
	\includegraphics[width=\columnwidth]{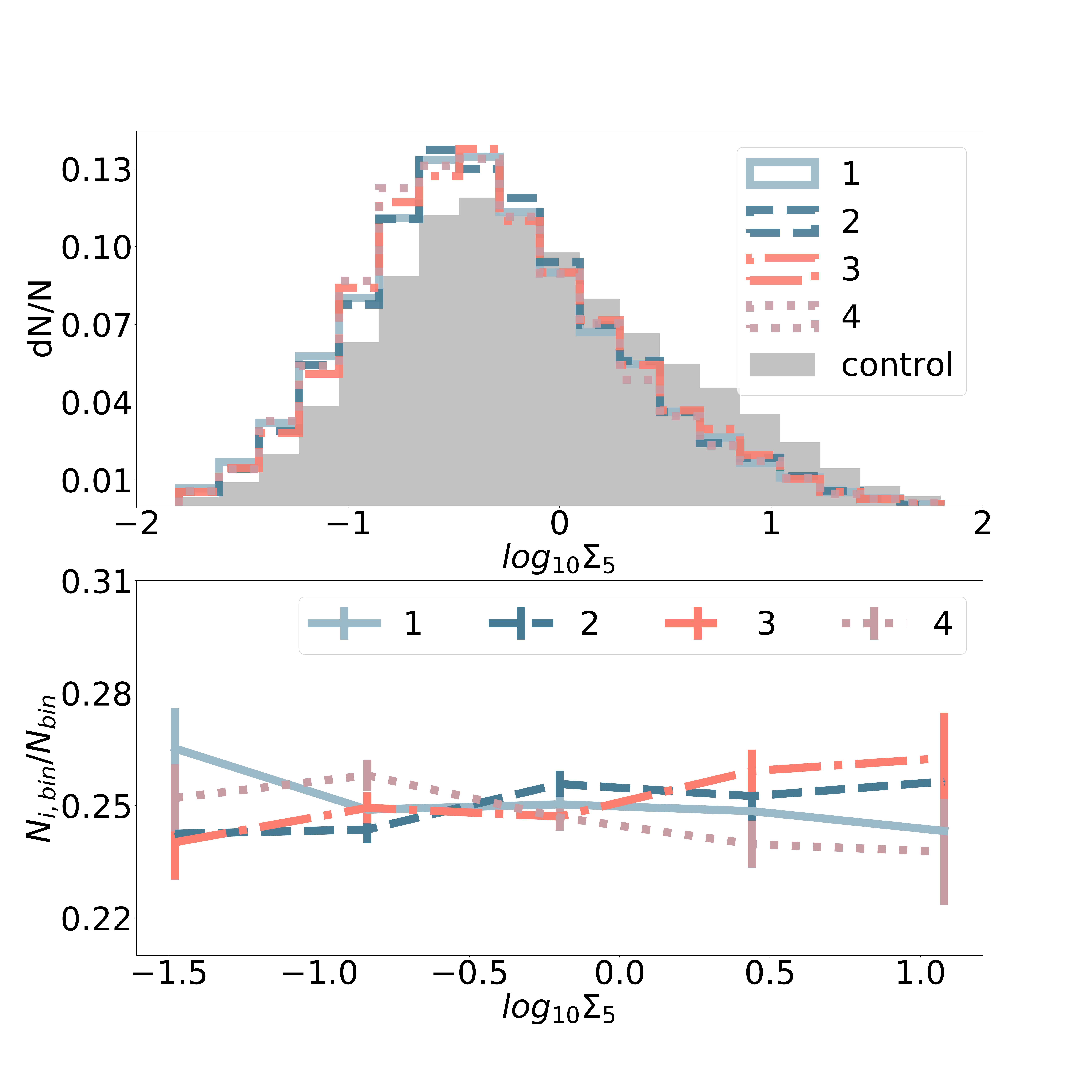}
    \caption{Top: Normalized distributions of the surface density in logarithmic scale, $\Sigma_5$, for the AGN subsamples and control sample (shadow).
    Bottom: Fraction of AGNs as a function of the surface density, $\Sigma_5$, in logarithmic scale. Subsamples ``1'', ``2'', ``3'', and ``4'' are represented by solid, dashed, dash-dotted and dotted lines, respectively.}
    \label{fig:perp_Hsigma5}
\end{figure}

For the sake of consistency, we perform the analysis of the previous section to the perpendicular subsamples with the aim to find an analogy in their behaviour regarding to the environment.
Then, in Fig. \ref{fig:perp_Hsigma5} (top panel) we show the $\Sigma_5$ distributions of AGNs from every subsample. 
For comparison, the local density distribution of the control sample, built matching the galaxy properties of all the AGNs, is also included. 
In this figure we cannot observe evident differences in the AGN subsample distributions.
Therefore, to quantify possible variations between the four subsamples we estimate the fraction of AGN galaxies belonging to every perpendicular bin, with respect to the total number of objects for a given bin of $\Sigma_5$.

In spite of the appreciable variations of the perpendicular AGN subsample host properties observed in the bottom panels of Figs. \ref{fig:sfr_gr_prop} and \ref{fig:dn_c_prop}, we can observe in Fig. \ref{fig:perp_Hsigma5} (bottom panel) that the fractions of galaxies belonging to every perpendicular bin, with respect to the total number of objects for a given bin of $\Sigma_5$, are comparable within the errorbars.
Subsample ``1'' presents a slightly higher fraction of galaxies at low values of $\Sigma_5$, within 1 $\sigma$ error. 
This tendency decreases from 27 per cent to 24 per cent, approximately, for the highest density. 
A similar trend, within the errorbars, is observed for galaxies from bin ``4''.
subsamples ``2'' and ``3'' have a lower fraction of galaxies at low local densities, showing a weak increasing of 1.4 per cent and 2.2 per cent, respectively, at the highest local density. 
These small detected variations present a good agreement with the results of \citet{Carter2001} and \citet{Amiri2019} who found that the fraction of galaxies with nuclear activity do not depend on the density of the environment. 

\begin{figure}
	\includegraphics[width=\columnwidth]{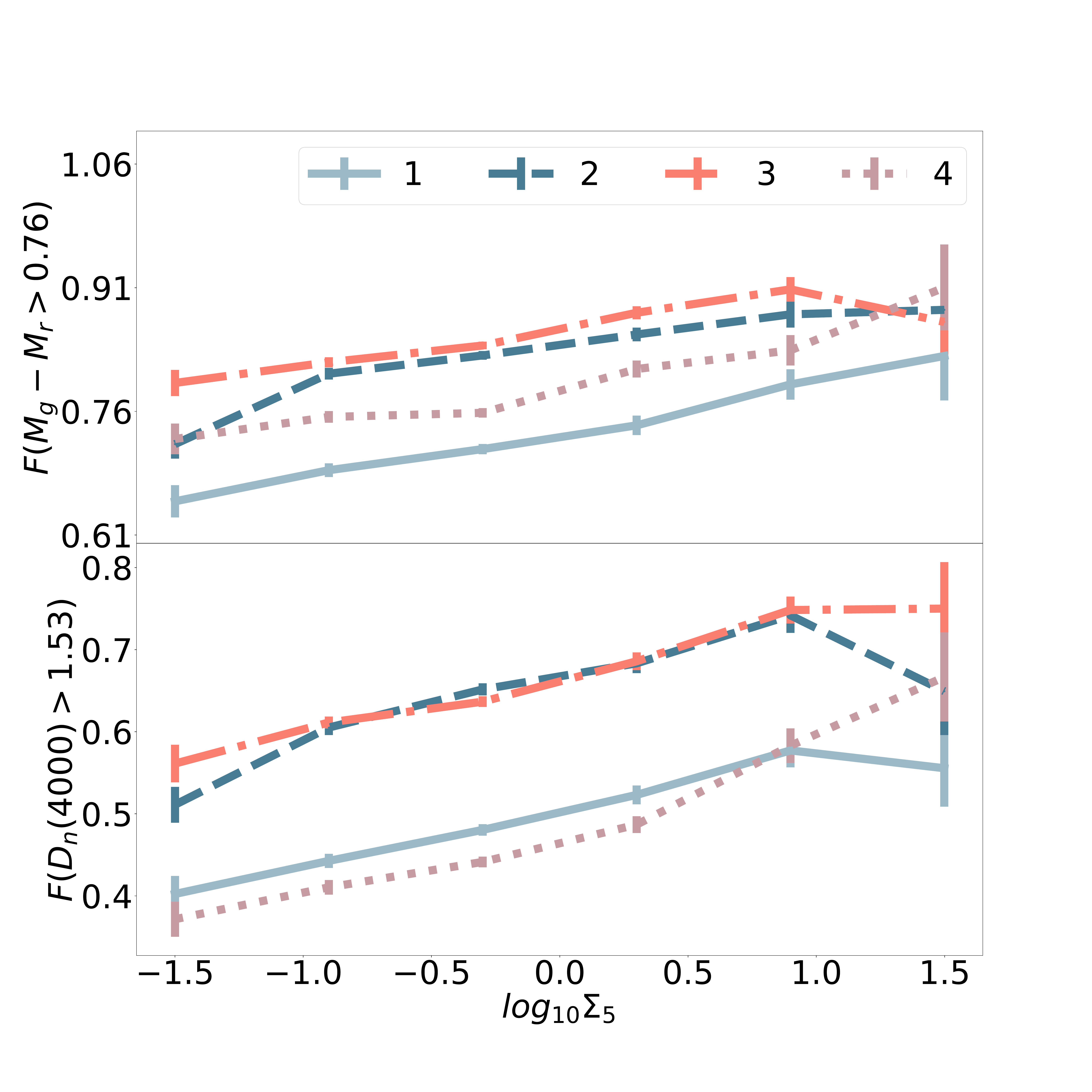}
    \caption{Fraction of red ($M_g - M_r > 0.76$) and old ($D_n(4000) > 1.53$) galaxies, as a function of the local density, for every perpendicular subsample. Subsamples ``1'', ``2'', ``3'' and ``4'' are represented by the solid, dashed, dash-dotted and dotted lines, respectively.}
    \label{fig:perp_Fprop-S5}
\end{figure}

In addition, Figure \ref{fig:perp_Fprop-S5} shows the fraction of galaxies, in every subsample, older and redder than the average values ($M_g - M_r > 0.76$ and $D_n(4000) > 1.53$) as a function of the local density. 
Here, we can observe an increasing fraction of red galaxies with the density environment, being more noticeable for galaxies belonging to bins ``1'', ``2'',  and ``4'', with an increment of $\sim 20$ per cent.
The tendency for galaxies from bin ``3'' is less sensitive to the environments.
The trends observed for the fraction of old galaxies suggest a major dependency with the environment than that from the galaxy colours for these subsamples, being consistent with the property distributions observed in Fig. \ref{fig:dn_c_prop}.
In addition, the trend for subsample ``1'' is similar to that for subsample ``4'', and the tendency from bins ``2'' and ``3'' are  comparable, except for high-density environments, where errorbars are significantly higher.

\begin{figure}
	\includegraphics[width=\columnwidth]{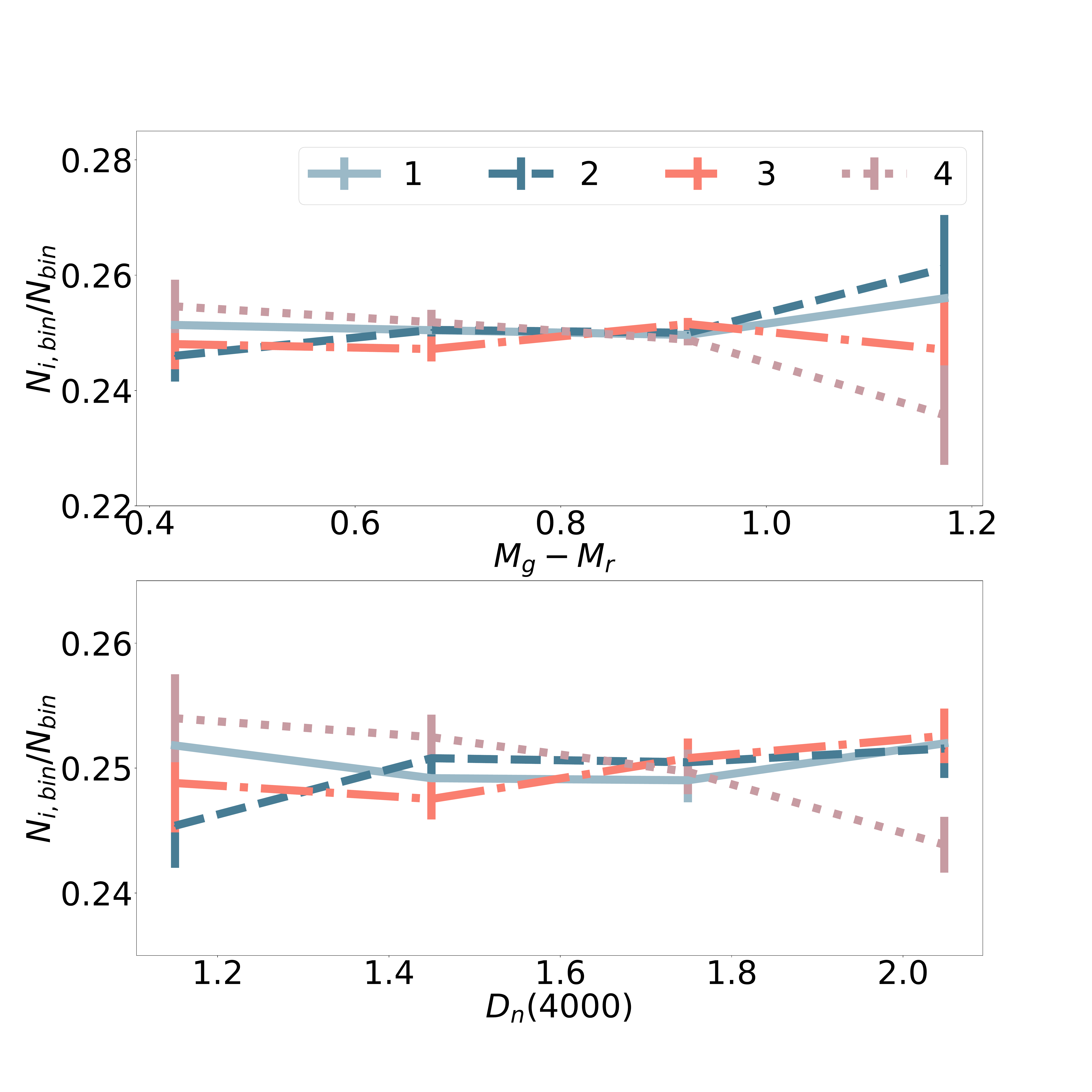}
    \caption{Fraction of the five closest neighbouring galaxies from AGNs as a function of the
    $M_g - M_r$ colour and the $D_n(4000)$ stellar age indicator. Subsamples ``1'', ``2'', ``3'', and ``4'' are represented by the solid, dashed, dash-dotted and dotted lines, respectively.}
    \label{fig:perp_H5tavecina}
\end{figure}

Regarding to the local galaxy neighbourhood,  we analyse the colours and stellar age of the five closest galaxies, as defined in the previous section. 
In Figure \ref{fig:perp_H5tavecina}, we show the fraction of galaxy neighbours of the subsamples with respect to the total number of objects for a given bin of $M_g-M_r$ and $D_n(4000)$. 
The trends for the colours of the surrounding galaxies of AGNs from all the subsamples are almost indistinguishable, except for redder colours where subsample ``2'' shows an excess of 1.1\% of neighbouring red galaxies and subsample ``4'' have a lack of 1.3\%, both respect to the mean values. 
The trends for the stellar age indicator of the five closest neighbouring galaxies in subsamples ``2'' and ``3'' are considerably similar. The $D_n(4000)$ indicator of surrounding galaxies of bin ``1'' has an approximately constant distribution.
On the other hand, the vicinity of AGNs from subsample ``4'' presents a deficit of 0.86\% of old galaxies with respect to the remaining subsamples.
From Table \ref{tab:muestras}, we observe that this last subsample, having the younger vicinity, corresponds to AGNs mainly classified as Composite galaxies with the highest fraction of spiral morphology.

\section{Summary and discussion}

We have performed a statistical analysis of the narrow emission-line AGNs derived from SDSS-DR12 data, in the redshift range $0.04 < z < 0.1$, studying the dependence of their host properties and environments with their position on the BPT diagram.

In the work of \citet{Kewley2006} the authors studied the host properties as a function of the distance from two empiric points, in the $\text{[O III]} / \text{H} \beta$  vs $\text{[O I]} / \text{H} \alpha$ diagnostic diagram, defined to determine a linear distance from the star-forming sequence for Seyfert and LINER galaxies, finding that the stellar population of both AGN samples become older at large distance from these points.

From our perspective, in this work we define AGN subsamples according to their position on the BPT diagram taking as references the Ka03 and Sc07 criteria, allowing a detailed analysis of the host properties and environment. Thus, the whole AGN sample was divided in parallel and perpendicular subsamples as we describe below:
Parallel subsamples were obtained by shifting a curve line parallel to the line criterion defined by Ka03, to separate AGN from star-forming galaxies.
The subsamples are named as ``a'',``b'',``c'', and ``d'', where ``a'' is closest to the  Ka03 criterion and ``d'' is the set of galaxies farthest from the Ka03 criterion.
Perpendicular subsamples were determined by moving the line proposed by Sc07 to separate LINER from Seyfert, from left top to right bottom  on the BPT diagram.
The subsamples are named ``1'',``2'',``3'', and ``4'', where ``1'' is the subsample lying closest to the top of the diagram.
The bins of all AGN subsamples have variable width in order to contain similar number of sources.

The main results can be summarised as follow:
\begin{itemize}
\item Parallel subsamples
\begin{itemize}
    \item The host galaxy property distributions show that colour, stellar age estimator, concentration index, and SFR have significant variations depending on their position on the BPT diagnostic diagram. Galaxies belonging to ``a'' subsample have higher values of SFR and are bluer and younger than galaxies from bin ``d'', while galaxies in between them present intermediate behaviour. Also, the concentration index parameter indicates that galaxies closer to the Ka03 line are less compact than the farthest ones, in agreement with the morphological tendency, estimated from Galaxy Zoo, shown in Table \ref{tab:muestras}. 
This finding is strongly related with percentage of AGN subclasses in each bin. The  ``a'' and ``b'' subsamples are almost entirely formed by Composite galaxies meaning that these sources have a mixing of HII regions and AGN emission.
The ``c'' subsample has similar proportion of LINER and Composite galaxies and ``d'' subsample is composite mainly for LINER and Seyfert AGN.
 Thus, when considering perpendicular distances from the tight sequence of star-forming galaxies it is possible to observe a clear progression on the properties, indicating the existence of a morphological evolution of the AGN host galaxies as they become more distant to the Ka03 criterion, in agreement with \cite{Kewley2006} and \citet{Zewdie2020}.
 
 \item Furthermore, we have studied the local environment of AGN host galaxies using the local density estimator parameter. 
We calculate the fraction of galaxies with respect to the total number of objects for each bin of $\Sigma_5$. Since the whole AGN galaxy sample do not follow the expected morphology-density relation as it is observed for the control sample of star-forming galaxies selected to match the AGN distributions of redshift, colour, concentration index, and stellar age, we perform the analysis for the selected subsamples.
Galaxies belonging to subsample ``a'' show an evident morphology-density relation, while AGNs in subsamples ``b'',``c'',  and ``d'' do not follow this relation strictly.
The result observed for galaxies from bin ``a'' can be predicted because this subsample is entirely formed by Composite galaxies where the relative contribution of the AGN is mixed with that from HII regions. 
However, when galaxies move away from the locus of the star-forming sequence, the nuclear activity becomes the more dominant source emission and the host galaxies tend to avoid to follow the morphology-density relation \citep{Miller2003,Padilla2010}.

\item Also, we analyse the dependence of the most evolved host galaxies with respect to the environment calculating the fraction of galaxies redder and older than average values ($M_g - M_r > 0.76$ and $D_n(4000) > 1.53$) as a function of the local density. It is possible to observe that the fraction of bulge-type galaxies, redder and older than the mean galaxy population, increases significantly from bin ``a'' to ``d'' as they move away from the Ka03 line.
It is noticeable that the fraction of redder galaxies increases slightly from low to high local density environments, being more evident for the subsample ``a''. The fractions of older galaxies are less sensitive to the density environment. 

\item The complementary study of the properties of the five AGN closest neighbouring galaxies brighter than $M_r=-20.5$, used to estimate $\Sigma_5$, shows that these galaxies present  similar fraction of both colours and stellar ages, with small variations between the subsamples with respect to the mean proportion.
The higher difference is shown by the colours of the neighbour galaxies of AGNs belonging the bin ``a'' reaching a maximum of  3$\sigma$, for the bluest colours,  regarding to the neighbours of AGNs from subsamples  ``c'' and ``d''.
Also, the fractions of red and old galaxies around AGNs from bin ``a'' are the lowest ($\sim $2.5\% respect to neighbour galaxies of AGNs from subsample ``d'' ).
\end{itemize}
\end{itemize}

\begin{itemize}
\item Perpendicular subsamples
\begin{itemize}

\item The normalized property distributions of the concentration index show a slightly excess of disc-type galaxies lying to subsample ``4'', consistent with the highest fraction of spiral galaxies from Table \ref{tab:muestras}. Furthermore, AGNs from subsample ``4'' show higher values of SFR than those from subsample ``1'' although galaxies in subsample ``1'' are the bluest and youngest and mainly classified as Seyfert. AGNs from bins ``2'' and ``3'' present intermediate galaxy colour and concentration index distributions.
The $D_n(4000)$ and SFR distributions for subsample ``2'' and ``3'' present a remarkable bimodality indicating the presence of two distinctive galaxy populations since these samples are formed by LINER and Composite galaxies. 
The fraction of LINER and elliptical galaxies reach the maximum fraction for subsample ``2''.
The host properties observed for galaxies in each subsample are directly related to the morphology and the percentage of AGN subclasses \citep{Kewley2006}.

\item Regarding the density environment, the subsamples ``1'' and ``4'' present a slightly higher fraction of galaxies at low values of $\Sigma_5$, decreasing for the highest density.
Subsamples ``2'' and ``3'' have a lower fraction of galaxies at low local densities, showing a weak enhancement at the highest local density. These small variations are of the order of the errors, although the host galaxies from each subsample have distinctive host properties, in agreement with the previous results of \citet{Carter2001} and \citet{Amiri2019}.
 
\item Related with the fraction of red and old galaxies with the density environment it is possible to observe a tendency to find higher fraction of these galaxies at higher densities. The $D_n(4000)$ parameter shows a major dependency with the environment than that from the galaxy colours for all the subsamples, with an increment of ~20 per cent for high-density environments.

\item From the study of the local galaxy neighbourhood of AGNs from perpendicular bins, the colour distributions of the five closest galaxies show that the fractions of these surrounding galaxies for a given bin of $M_g-M_r$ are indistinguishable for all the subsamples, except for redder colours where the errorbars are large.
The distribution of the stellar age of the five closest galaxies in subsample “1” is approximately constant. Surrounding galaxies of AGNs from bins “2” and “3” are pretty similar within the errors. Only the environments of subsample “4” present a lack of old galaxies with respect to the remaining subsamples, although this is lesser than 1\% .

\end{itemize}
\end{itemize}

Summarizing, we have found a clear relationship between the position of the AGNs across the BPT diagram and their host galaxy properties considering both parallel and perpendicular subsamples.
A strong evolution is observed for host AGNs from parallel subsamples which are redder, older, and with more prominent bulge-type morphology as they drift away from the tight star-forming sequence. Host galaxies from perpendicular subsamples present features according to the proportion of AGN subclasses and visual morphological classification. However, even when the AGN characteristics of every subsample are very distinctive, the environment does not seem to reveal this fact.
The density environment, and the properties of the AGN galaxy neighbourhood of each subsample appear be associated with the AGN subclasses where the subsamples containing higher fraction of Composite and Spiral galaxies are able to show a more evident morphology-density relation than those mainly formed by "pure" AGNs, Seyferts, and LINERs, in agreement with previous works 
\citep{Miller2003,Padilla2010,Coldwell2014,Coldwell2017}.

Observational evidence suggests that most massive galaxies harbour a central BH \citep{Tremaine2002, Marconi2003}, and even  some dwarf galaxies may contain less massive intermediate-mass black holes \citep{Mezcua2018}; however, not all these galaxies are active. Several physical processes involved in the galaxy evolution,
such as interactions \citep{Alonso07}, disc instabilities producing bars  \citep{Alonso18},  major and minor mergers \citep{Secrest2020}, etc. 
have been proposed as those responsible of triggering the AGN activity. 
 Thus, these mechanisms can occur in different stage of the galaxy life being able to reactivate a dormant supermassive black hole \citep{Coldwell2009}. 
 In this scenario younger host galaxies with high gas content, such as AGNs with spectral mixing contribution from HII regions, could have enough fuel to feed the black hole in their early stage regardless of the environmental condition. 
However the more evolved host galaxies, where AGNs become the dominant emission source,  could depend of a trigger mechanism occurring when the intergalactic medium has high content of gas and a major likelihood of a higher merger rate providing more suitable conditions for the central black hole feeding.

\section*{Acknowledgements}

We thank the referees, for providing us with helpful comments and suggestions that improved this paper.
This work was supported in part by the Consejo Nacional de 
Investigaciones Cient\'ificas y T\'ecnicas de la Rep\'ublica Argentina 
(CONICET) and the Consejo Nacional de Investigaciones Cient\'ificas, T\'ecnicas y de Creaci\'on Art\'istica de la 
Universidad Nacional de San Juan (CICITCA).

Funding for SDSS-III has been provided by the Alfred P. Sloan Foundation, the Participating Institutions, 
the National Science Foundation, and the U.S. Department of Energy Office of Science.
The SDSS-III web site is \emph{http://www.sdss3.org/}.
SDSS-III is managed by the Astrophysical Research Consortium for the Participating Institutions of the SDSS-III 
Collaboration including the University of Arizona, the Brazilian Participation Group, Brookhaven National Laboratory, 
Carnegie Mellon University, University of Florida, the French Participation Group, the German Participation Group, 
Harvard University, the Instituto de Astrofisica de Canarias, the Michigan State/Notre Dame/JINA Participation Group, Johns Hopkins University, Lawrence Berkeley National Laboratory, Max Planck Institute for Astrophysics, Max Planck 
Institute for Extraterrestrial Physics, New Mexico State University, New York University, Ohio State University, 
Pennsylvania State University, University of Portsmouth, Princeton University, the Spanish Participation Group, 
University of Tokyo, University of Utah, Vanderbilt University, University of Virginia, University of Washington, and Yale University.

\section*{DATA AVAILABILITY}
The data underlying this article will be shared on reasonable request to the corresponding author.


\bibliographystyle{mnras}
\bibliography{biblio}

% Don't change these lines
\bsp	% typesetting comment
\label{lastpage}
\end{document}